\documentstyle[preprint,prl,aps]{revtex}
\input{epsf}

\begin{document}
\draft
\title{Absence of bimodal peak spacing distribution 
in the Coulomb blockade regime}

\author
{Richard Berkovits}

\address{
The Minerva Center for the Physics of Mesoscopics, Fractals and Neural 
Networks,\\ Department of Physics, Bar-Ilan University,
Ramat-Gan 52900, Israel}

\date{\today}
\maketitle

\begin{abstract}
Using exact diagonalization numerical methods, as well as analytical arguments,
we show that for the typical electron densities in chaotic and disordered 
dots the peak spacing distribution is not bimodal, but rather
Gaussian. This is in agreement with the 
experimental observations. We attribute this behavior to the tendency of 
an even number of electrons to gain on-site interaction energy by removing the
spin degeneracy.
Thus, the dot is predicted to show a non trivial electron number
dependent spin polarization. Experimental test of this hypothesis based on 
the spin polarization measurements are proposed.
\end{abstract}

The distribution of the addition spectrum of chaotic and disordered
quantum dots measured in recent
experiments \cite {sba,shw,charlie}
seem to contradict the predictions of the orthodox constant
interaction model \cite{Kastner,Meirav,Ashoori,McEuen}.
The most striking feature in the
addition spectrum distribution is the absence of any signs of a bimodal
structure corresponding to the two spin states of the tunneling electron.
In the constant interaction (CI) approximation, the
ground state energy of a quantum dot populated by $N$ electrons is given
by $E_N= {e^2 N^2}/{2C}+\sum_{i=1}^{N_{\uparrow}}\eta _i + 
\sum_{i=1}^{N_{\downarrow}}\eta _i$ 
where $C$ is the dot's
constant (or slowly varying) capacitance, $\eta _i$ are the single
particle energies, and $N_{\uparrow}$ ($N_{\downarrow}$) 
is the number of electrons
for which the spin is in the up (down) state ($N=N_{\uparrow}+N_{\downarrow}$).
In the absence of a magnetic field, for an even number of electrons
$N_{\uparrow}=N_{\downarrow}=N/2$ while for an odd $N$, $N_{\uparrow}=(N+1)/2$
and $N_{\downarrow}=(N-1)/2$. Thus, the change in the chemical potential
needed to add a particle, $\Delta_2 = \mu_N - \mu_{N-1} =
E_{N} - 2E_{N-1} + E_{N-2}$, is equal to $e^2/C$ for $N$ even,
while $\Delta_2 = e^2/C + \eta_{(N+1)/2} - \eta_{(N-1)/2}$ for
$N$ odd. Assuming that the single electron level spacings follow
the random matrix theory (RMT) predictions, the distribution of the
spacings should be, in the absence of a magnetic field,
\begin{eqnarray}
P_{CI+RMT}(s)={{1}\over{2}} 
\left(\delta(s) + \pi s \exp(-\pi s^2) \right),
\label{ps}
\end{eqnarray}
where $s=(\Delta_2 - e^2/C)/2 \Delta$, 
$\Delta=\langle \eta_i-\eta_{i-1} \rangle$ is the averaged single level
spacing and $\langle \ldots \rangle$ denotes an average over realizations
or levels. As we have mentioned, none of the experimental distributions
shows even the slightest sign of any bimodal distribution.

When the interactions are treated beyond the CI model, different distributions
emerge. In Ref. \cite{Blanter}, the influence of the interactions
on the distribution within the Hartree approximation
was considered. For strong interactions, this, roughly speaking, leads to
a distribution composed of two uncorrelated Wigner distributions resulting
in the spin resolved (SR) RMT distribution
(see \cite{charlie,Metha})
which is plotted in Fig. \ref{fig.2}a.
This distribution begins abruptly ($P(s<0)=0$ while $P(s=0)=1/2$)
and is strongly asymmetric. All the experimental distributions are 
nevertheless more or less Gaussian with only weak asymmetries in the tails
\cite {sba,shw,charlie}.

Thus, none of the above models can satisfactorily reproduce the experimental
distributions. We therefore use an exact diagonalization study
to investigate the role of spin in determining the addition spectrum of 
a disordered tight-binding Hamiltonian with on-site and long range
interactions. Although we pay the price of handling only small systems,
we avoid the uncertainty of approximations and fully include the effects of 
electron correlations. 
In similar studies of symmetric dots, atoms and nuclei 
\cite{Ashoori,McEuen,Kouw} the spatial symmetry
of the confining potential plays an important role and leads to a
strong dependence on the particle numbers via the shell structure 
(magic numbers) and
a non-trivial total spin due to Hund rules. Because of the chaotic nature
of the dots \cite {sba,shw,charlie} we do not expect such dependence
on the electron number to play an important role, and therefore 
the study of a small number of electrons is still useful in understanding
the properties of dots which are populated by an order of magnitude more 
electrons. Both components of the Coulomb interaction on a lattice, 
i.e., the long range part 
and the on-site 
interaction must be considered since without the long range component of the
Coulomb interaction
the classical capacitance behavior for the average spacing will
not be reproduced, and for electrons with spin on a lattice a Coulomb 
interaction must have an on-site component. 

Three different regions of behavior emerge. The weak interaction limit,
which corresponds to an on-site interaction parameter smaller
than a fifth of the bandwidth, is characterized by a strong even-odd
asymmetry of the addition spectrum. As discussed below this region
correspond to values of $r_s< 0.1-0.3$ (where $r_s$ is the ratio of the
inter-particle Coulomb energy to the kinetic energy). An intermediate
region corresponding to $0.1-0.3<r_s<1$, where there is no sign of
the spin in the $\Delta_2$ distribution, the distribution tends towards
Gaussian, and the total spin of an even number of electrons in the
dot is likely to be partially polarized, while the spin of an odd number of
electrons is hardly affected and is equal to $\hbar/2$. 
For the strong interaction regime
$r_s>1$ the distribution is a Gaussian of a width proportional to $e^2/C$,
and the spin polarization of the dot shows a complicated behavior.

Since all experiments are in the $r_s\sim 1$ regime \cite {sba,charlie}
these results clearly are consistent with no bimodal distribution 
being observed in the experiment. As we shall see, the unimode nature of the 
distribution can be understood by treating more clearly the ground state spin
degeneracy.

We calculate the ground state energies 
for a system of interacting electrons modeled by 
a tight-binding 2D Hamiltonian given by:
\begin{eqnarray}
H= \sum_{k,j,s} \epsilon_{k,j} a_{k,j,s}^{\dag} a_{k,j,s} - V \sum_{k,j,s}
(a_{k,j+1,s}^{\dag} a_{k,j,s} + a_{k+1,j,s}^{\dag} a_{k,j,s} + h.c)
+ H_{int},
\label{hamil}
\end{eqnarray}
where  
$\epsilon_{k,j}$ is the energy of a site ($k,j$), chosen 
randomly between $-W/2$ and $W/2$ with uniform probability, $V$
is a constant hopping matrix element, and
$s=\uparrow,\downarrow$ is the spin. The interaction
Hamiltonian is given by:
\begin{equation}
H_{int} = U' \sum_{k,j} a_{k,j,\uparrow}^{\dag} a_{k,j,\uparrow}
a_{k,j,\downarrow}^{\dag} a_{k,j,\downarrow} +
U  \sum_{k,j>l,p;s,s'} {{a_{k,j,s}^{\dag} a_{k,j,s}
a_{l,p,s'}^{\dag} a_{l,p,s'}} \over 
{|\vec r_{k,j} - \vec r_{l,p}|/b}}
\label{hamil2}
\end{equation}
where $U'$ is the on-site interaction constant and $U=e^2/b$ 
is the long range component of the Coulomb interaction constant where $b$ 
is a lattice constant. 

We consider a $4 \times 3$ dot with $M=12$ sites and 
up to $N=9$ electrons. The size of the many-body Hilbert space is 
$m = (_N^{2M})$, thus for $M=12$ and $N=9$ we end up with $m=1,307,504$. 
One can use the fact that in the site representation the Hamiltonian
matrix has no off-diagonal terms which couple states of different $S_z$
(where $S_z$ is the component of the total spin in the $\hat z$ direction) 
to diagonalize 
each block with a given $S_z$ separately. The size of each block is
$(_{N_{\uparrow}}^M)\times(_{N_{\downarrow}}^M)$ which for the largest case
is equal to $392,040$. After the minimal eigenvalue for each block is obtained
the global minimal eigenvalue is found. Since the system has a $2S+1$ 
ground state degeneracy
(where $S$ is the total ground state spin), for all the cases in which
the total ground state spin $S \neq 0$ all the blocks corresponding to
$S_z = -S, -S+1 \ldots S$ 
have the same minimal eigenvalue. This gives an excellent
check for the accuracy of the diagonalization procedure. Therefore,
both the ground state energy and the value of $S$ are found by the exact
diagonalization.

The strength of the long range component of the
Coulomb interactions, $U$, is varied between $0-6V$. 
For the results presented here, an on-site coupling $U'=(10/3)U$ was chosen, 
corresponding to Hubbard's
calculation of the ratio of $U'$ to $U$ for weakly overlapping hydrogen
like wave-functions \cite{Hubbard}. As will be discussed later 
the exact ratio of $U'$ to $U$ does not play a crucial role in determining
the main results presented here.
The disorder strength is set to $W=8V$
in order to assure perfect RMT behavior
for the non-interacting case. We also present some data pertaining to
the $W=3V$ case, 
where only the lower levels (up to $N=6$) which show RMT behavior
were considered in  order to gain some insight into the role of disorder.
For each value of $U$, the results for each value of $N$ are averaged over
$500$ different realizations of disorder.

The results of the tight-binding
calculations at a given interaction strength $U$ can then be compared to the 
experimental quantum dot density parameters. The ratio of the
average inter-particle Coulomb interaction and the Fermi energy
$r_s=1/\sqrt{\pi n} a_B$ (where $n$ is the electronic density and
$a_B$ is the Bohr radius) corresponds to
$r_s\sim\sqrt{\pi/2}(U/4V)$ for $N=6$, $M=12$. For all experimental setups
$r_s\sim 1$ \cite {sba,charlie}, resulting in $U \sim 4V$.

\begin{figure}
\centerline{\epsfxsize = 4in \epsffile{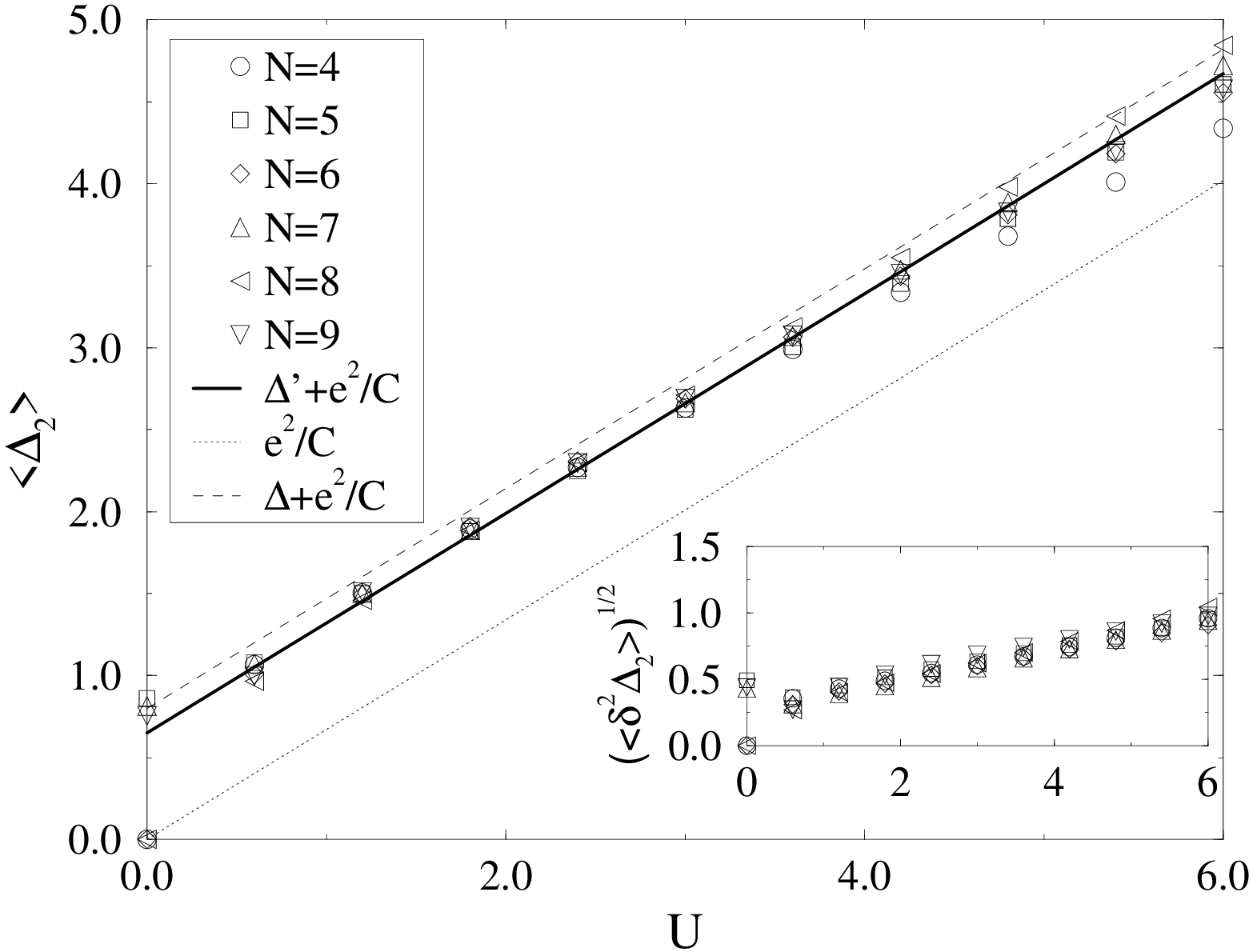}}
\caption {The value of $\protect \langle \Delta_2 \rangle$ as function
of $U$ for a $\protect 4 \times 3$ lattice and a different number of electrons
$N$. All values are in units of the hopping matrix element $V$. Inset: the
fluctuations .
\label{fig.1}}
\end{figure}

\begin{figure}
\vspace{-1.5cm}
\centerline{\epsfxsize = 3in \epsffile{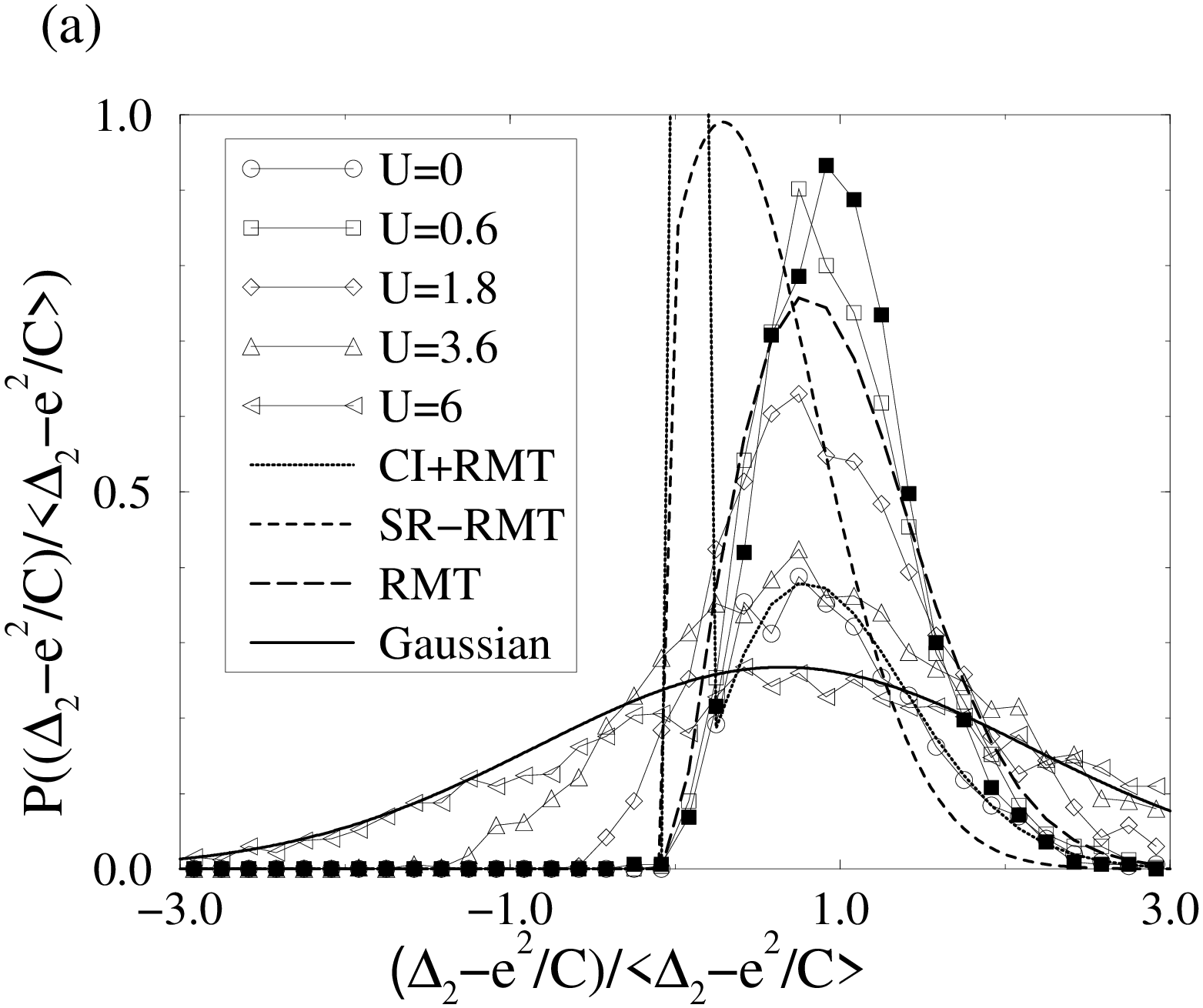}}
\vspace{-1cm}
\centerline{\epsfxsize = 3in \epsffile{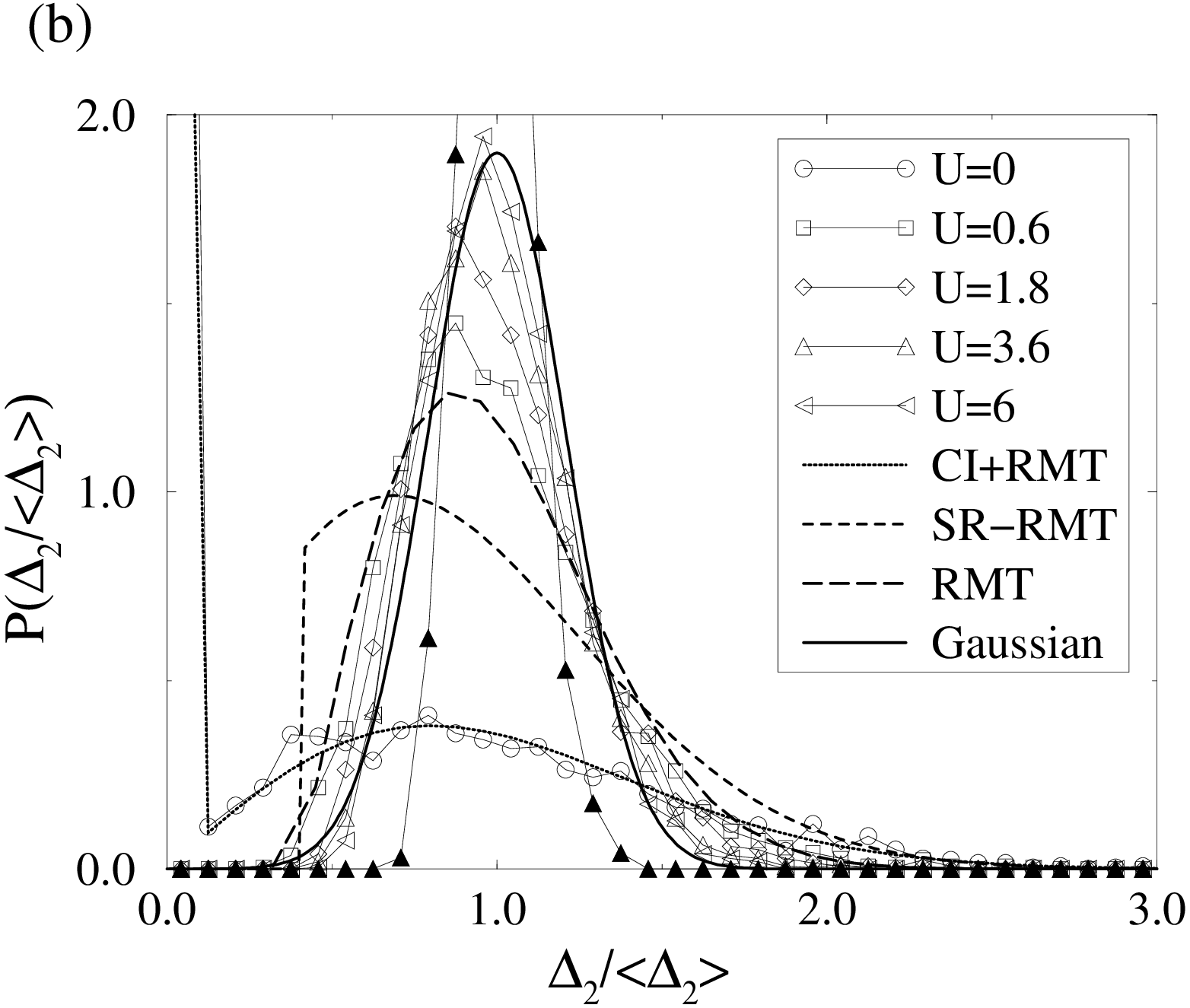}}
\caption {The probability distribution of $\Delta_2$ for different
values of the interaction $U$, drawn on the scale
of (a) $\Delta_2 - e^2/C$ and of (b) $\Delta_2$. The heavy lines correspond
to Eq. (\protect \ref{ps}) with $e^2/C=0$,
SR-RMT, Eq. (\protect \ref{pr}) and to 
a Gaussian with a variance of $0.2e^2/C$. Open symbols correspond to
$W=8V$ while filled symbols to $W=3V$. 
\label{fig.2}}
\end{figure}

The average value $\langle \Delta_2 \rangle$ 
is presented in Fig. \ref{fig.1}. In the
non interacting case ($U=0$), $\langle \Delta_2 \rangle = 0$
for even electron numbers, while for odd numbers 
$\langle \Delta_2 \rangle = \Delta$. Even for weak interactions
($U=0.6V$ corresponding to $r_s \sim 0.2$) there is no remnant of the
even-odd asymmetry in $\langle \Delta_2 \rangle$. Moreover, up to $U\sim4$,
for all values of $N$, $\langle \Delta_2 \rangle = e^2/C +\Delta'$, where
the capacitance $e^2/C = 0.67U$ was calculated through the 
random phase approximation 
(for details see Ref. \cite{ba}) and has no adjustable parameters, and
$\Delta' = 0.81 \Delta$ is obtained from a fit. Since this behavior appears
already at weak interactions and the influence of the long range component of 
the Coulomb 
interaction is well described by
the capacitance up to $U\sim 4V$ ($r_s=1$)
it is natural to concentrate on the role of the on-site interactions. In
first order perturbation theory for the on-site interaction strength, 
where the single electron eigenfunctions
are assumed to follow the random vector model (RVM), 
it has been shown \cite{prus}
that  $\langle \Delta_2 \rangle = e^2/C + 3U'/(M+2)$ 
for an even number of electrons
and  $\langle \Delta_2 \rangle = e^2/C + \Delta - 2U'/(M+2)$ for the odd case.
Thus, the even and odd values of $\langle \Delta_2 \rangle$ will coincide at 
$U_c'= (M+2)\Delta / 5 \sim 8V/5$, which corresponds here to $U=0.5V$, in good 
agreement with the numerical results. We have checked that this value of
$U_c'$ holds for different ratios between $U$ and $U'$. Thus, even
for the smallest conceivable ratio ($U'=U$, since $U'<U$ represents a locally 
attractive interaction) the even-odd asymmetry will disappear
for $r_s>0.4$, while a more reasonable estimation of the ratio will
yield the values quoted in the introduction. 
Above this value of interaction it is
not possible to continue to use this first order perturbation, but it is
reasonable to assume that the excess value of $\langle \Delta_2 \rangle$ above 
$e^2/C$ will be of order $\Delta'=3U_c'/(M+2)$, i.e., $\Delta'\sim 0.6 \Delta$,
which is not far from the numerical result. Above $U\sim 4V$ ($r_s=1$) short
range correlations in the electronic density appear and the random phase 
approximation is no longer valid, as discussed in Refs. \cite{sba,ba}.

The fluctuations $\langle \delta^2 \Delta_2 \rangle
= \langle (\Delta_2)^2 \rangle - \langle \Delta_2 \rangle^2$ 
are portrayed in the inset of Fig. \ref{fig.1}. As with the average, 
the even-odd
asymmetry disappears at $U=0.6V$ and a gradual enhancement 
of the fluctuations as a
function of the interaction strength is seen. 
The full distribution of $\Delta_2$
obtained for all values of $N$ ($N=4 \ldots 9$)
is shown in Fig. \ref{fig.2}. In Fig. \ref{fig.2}a the distribution is
plotted as function of $\Delta_2 - e^2/C$, which in a sense captures the
distribution of the ``single electron level'' spacings. 
In the non-interacting case
the CI+RMT distribution (Eq. \ref{ps}) fits perfectly. Even for weak 
interactions, $U=0.6V$ ($U'=2V$), the bimodal distribution is wiped away.
Neither does the SR-RMT distribution\cite{charlie,Metha} fit. The best fit 
for weak interactions is
obtained from the usual RMT distribution
\begin{eqnarray}
P_{RMT}(s)=(\pi s/2) \exp(-\pi s^2/4)
\label{pr}
\end{eqnarray}
where $s = (\Delta_2 - e^2/C) / \langle \Delta_2 - e^2/C \rangle$. 
This feature does not seem to depend much on the disorder, as can be seen
form the $W=3V$,$U=0.6V$ curve in Fig. \ref{fig.2}a.
As the interaction strength
increases, the distribution becomes wider, less asymmetric,
and from the fact that a considerable weight of the distribution is at 
negative values it becomes clear that no constant interaction model is able
to describe it. For the region of $r_s \sim 1$ ($U \sim 4V$)
a better fit of the distribution to a Gaussian is obtained.
The same results plotted as function of $\Delta_2/\langle \Delta_2 
\rangle$ are shown in  Fig. \ref{fig.2}b. A clear crossover from the RMT
like distribution to a Gaussian distribution is seen as the interaction
is increased, although its width does seem to depend on the disorder
(see the $W=3V$,$U=3.6V$ curve in Fig. \ref{fig.2}b).

\begin{figure}
\vspace{-1.5cm}
\centerline{\epsfxsize = 3in \epsffile{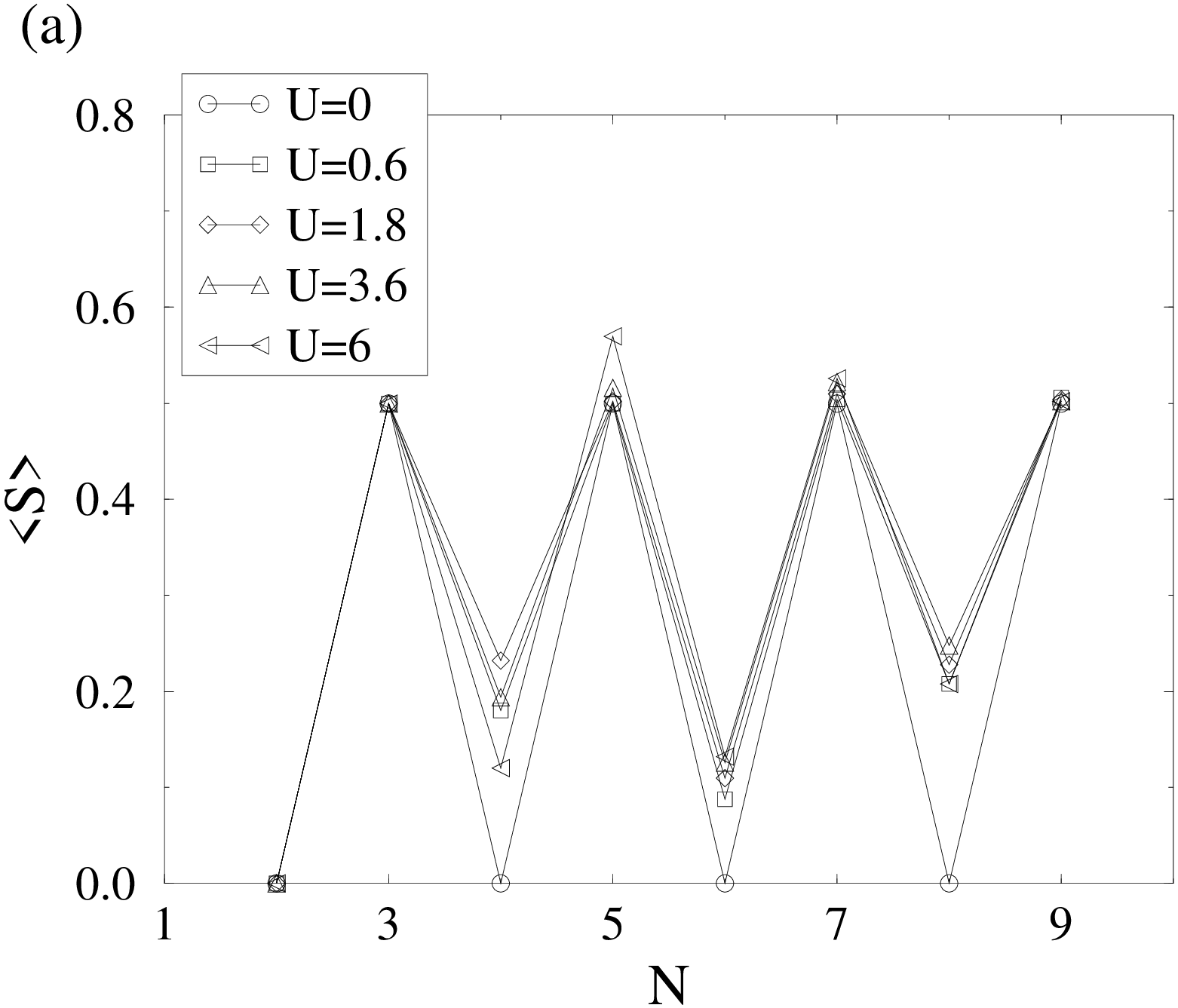}}
\vspace{-1cm}
\centerline{\epsfxsize = 3in \epsffile{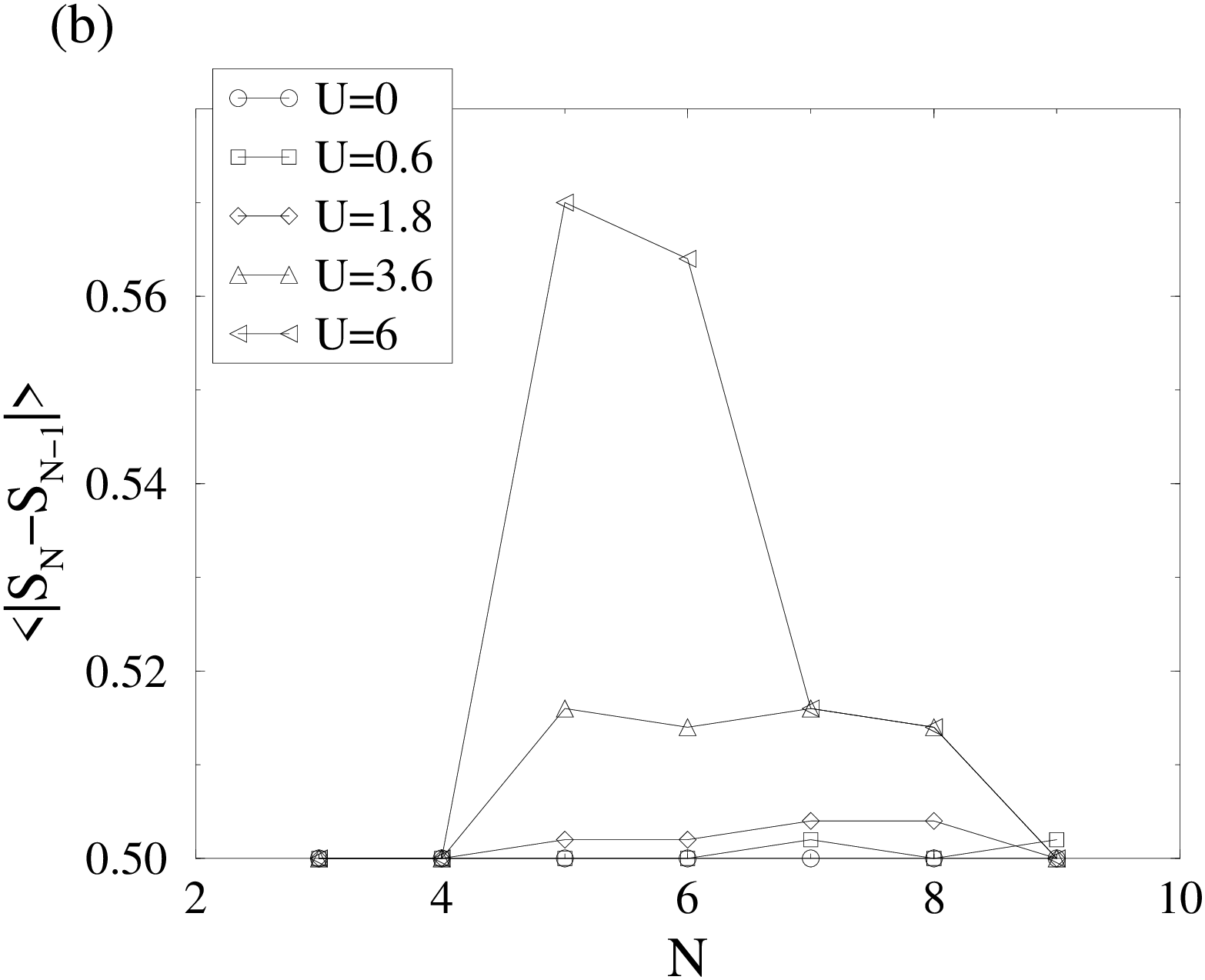}}
\caption {(a) The ground state average spin, and  (b) the average of the
absolute value of the change in the ground state spin as an electron is added
to the dot, as function of the number electrons
for different interaction strength.
\label{fig.3}}
\end{figure}

In all experiments a distribution which is Gaussian (up to deviations
in the tail) is seen \cite {sba,shw,charlie}. 
Since the experiments are performed in the region of
$r_s\sim 1$ this is in good agreement with our results. 
Leaving aside the width of the distribution for the moment, we need to
understand better the unexpected disappearance of the bimodal distribution
for intermediate coupling. We can gain insight into this
by studying the ground state polarization of the dot.
In the non-interacting case we expect the total spin of the dot to be
$S=0$ for an even number of electrons and $S=1/2$ for an odd number.
For weak interactions, using first order perturbation in the on-site
interaction strength with the RVM one finds that on the average
the system gains $3U'/(M+2)$
interaction energy by flipping one spin, while it loses 
$\eta_{N/2+1} - \eta_{N/2}$ kinetic energy for an even number of electrons,
compared to the gain of $4U'/(M+2)$ interaction energy and the loss of
$\eta_{(N+3)/2} - \eta_{(N-1)/2}$ kinetic energy for an odd number.
The dot thus will flip a spin if the gain in interaction energy will
be larger than the loss of kinetic energy \cite{prus}.
The probability of finding two consecutive small single electron level spacings
is much smaller than the probability of finding one small level spacing,
therefore we expect a finite probability for an even number of electrons
to be in a $S=1$ state, while a much lower probability to find
an odd number of electrons at $S=3/2$ is expected. Indeed in Fig. \ref{fig.3}a
it can be seen that while the average spin of the ground state for an odd 
number of electrons is $\langle S \rangle \sim 1/2$, except for strong 
interactions $U>4V$ ($r_s>1$), the average spin for the even case
$\langle S \rangle > 0$ also for rather weak interactions. 
Since both the interaction energy and the kinetic energy scale
as $1/M$ we expect this behavior to hold also for large systems.
Thus, in the
$r_s<1$ region we expect to see a significant number of $S=1$ states for
an even number of electrons and almost no $S=3/2$ states for an odd
number. Higher spin polarizations seem to be rather rare, 
although we encountered
two realizations with $S=2$ for $N=8$ and $U=1.8V$. Once interactions are 
stronger the even odd asymmetry in the ground state spin 
polarizability is less 
pronounced. This is illustrated in Fig. \ref{fig.3}b, where
$\langle |S_N - S_{N-1}| \rangle$ is plotted. For any non-correlated behavior
$|S_N - S_{N-1}| = 1/2$, while for correlated behavior adding an electron to
the dot may flip the spin of other electrons already in the dot and even lead
to spin blockade\cite{whk}.
A clear indication for the appearance of a correlated state is seen 
for $r_s>1$. 

Recently, Stopa \cite{stopa} has suggested that spin polarization in a chaotic
dot may appear due to scar states. According to this scenario the scar
state will be populated first by an $\uparrow$ electron, then several
other states of higher kinetic energy will be populated in the regular sequence
and eventually a $\downarrow$ will populate the scar state. This scenario
leads to a shift by a $1/2$ of $S_N$ between populating the scar state with 
the $\uparrow$ electron and the $\downarrow$ electron.
This is a distinct spin polarization pattern from the ones previously
described. The combined effect of scar states and regular chaotic or disordered
states considered in this letter
is an interesting question now under investigation.
A very recent paper \cite{vlm} has suggested
that the absence of the bimodal distribution is due to the deformation 
of the confining potential of the dot as electrons are added \cite{hhw}.
Again such a scenario will lead to a different spin polarization pattern 
than the previous ones, i.e., it will be equivalent to the non-interacting one.
The measurement of the dot's 
spin polarization is feasible\cite{ch}, 
probably via measuring the magnetic field dependence of the differential
conductance \cite{ralph}, from which the change in dot's spin may be 
observed.
As shown in this letter, by studying the change in the spin polarization
it is possible to clarify whether the underlying physics of the dot
corresponds to the correlated regime, the intermediate regime, 
a deformable potential, or maybe 
carries the signature of scar states. 

There remains a question regarding the width of the distribution. 
While earlier experiments show a width comparable to $0.15-0.10 e^2/C$
\cite {sba,shw}, a recent experiment\cite{charlie} yields a width 
of $0.05 e^2/C \sim \Delta$. Our simulations show 
$\sqrt{\langle \delta^2 \Delta_2 \rangle} \sim 0.2 e^2/C$ for $W=8V$ and
$\sqrt{\langle \delta^2 \Delta_2 \rangle} \sim 0.1 e^2/C$ for $W=3V$, 
consistent with previous results for spinless electron of 
$\sqrt{\langle \delta^2 \Delta_2 \rangle} \sim 0.1-0.2 e^2/C$, and with
the earlier experiments\cite {sba,shw}. Nevertheless, for the numerical
study at $U=4V$ ($r_s=1$), $0.2 e^2/C \sim \Delta$, which does not
contradict Ref. \cite{charlie}. It is clear that for $r_s \ll 1$, 
$\sqrt{\langle \delta^2 \Delta_2 \rangle} \propto \Delta$, while
for $r_s \gg 1$,
$\sqrt{\langle \delta^2 \Delta_2 \rangle} \propto e^2/C$. Whether
the behavior of the fluctuations in the above model
at $r_s\sim 1$
is determined by $\Delta$ or $e^2/C$, and the role of disorder,
could be determined only by
a careful finite size study, since the change in size will change
the ration of $\Delta$ to $e^2/C$ and enable the determination of the
relevant scale.

In conclusion, the on-site interaction is responsible of the removal of the
even-odd asymmetry in the addition spectrum distribution. These interactions
lead to a ground state spin polarizability of the dot. The dependence of
the spin polarizability on the number of electrons may be used as a 
sensitive tool to determine the relevant physics in the dot.

Many useful discussions on the addition spectrum of quantum dots with  
B. L. Altshuler, A. Auerbach, C. M. Marcus, A. D. Mirlin, O. Prus and U. Sivan
are gratefully acknowledged.
I would like to thank The Israel Science
Foundations Centers of
Excellence Program for financial support.

\end{document}